\documentclass[twocolumn,amsmath,floatfix,amssymb,aps,prb,showpacs,letter, superscriptaddress]{revtex4-1}
\usepackage[pdftex]{graphicx}
\usepackage{graphicx} % Include figure files

\usepackage{amssymb,amsmath}
\usepackage{dcolumn}
\usepackage{bm}
\usepackage{color,soul}
\begin{document}

\title{The elevated Curie temperature and half-metallicity \\
in the ferromagnetic semiconductor La$_{x}$Eu$_{1-x}$O}

\author{Pedro M.~S.~Monteiro}
\email{pmdsm2@cam.ac.uk}
\affiliation{Cavendish Laboratory, Physics Department, University of Cambridge, Cambridge CB3 0HE, United Kingdom}

\author{Peter~J.~Baker}
\affiliation{ISIS Facility, STFC Rutherford Appleton Laboratory, Harwell Science and Innovation Campus, Oxon, OX11 0QX, United Kingdom}

\author{Nicholas~D.~M.~Hine}
\affiliation{Cavendish Laboratory, Physics Department, University of Cambridge, Cambridge CB3 0HE, United Kingdom}
\affiliation{Department of Physics, University of Warwick, Coventry, CV4 7AL, United Kingdom}

\author{Nina-J. Steinke}
\affiliation{ISIS Facility, STFC Rutherford Appleton Laboratory, Harwell Science and Innovation Campus, Oxon, OX11 0QX, United Kingdom}

\author{Adrian~Ionescu}
\affiliation{Cavendish Laboratory, Physics Department, University of Cambridge, Cambridge CB3 0HE, United Kingdom}
\author{Joshaniel~F.~K.~Cooper}
\affiliation{ISIS Facility, STFC Rutherford Appleton Laboratory, Harwell Science and Innovation Campus, Oxon, OX11 0QX, United Kingdom}

\author{Crispin~H.~W.~Barnes}
\affiliation{Cavendish Laboratory, Physics Department, University of Cambridge, Cambridge CB3 0HE, United Kingdom}

\author{Christian J. Kinane}
\affiliation{ISIS Facility, STFC Rutherford Appleton Laboratory, Harwell Science and Innovation Campus, Oxon, OX11 0QX, United Kingdom}

\author{Zaher~Salman}
\affiliation{Laboratory for Muon-Spin Spectroscopy, Paul Scherrer Institut, CH-5232 Villigen, Switzerland}

\author{Andrew~R.~Wildes}
\affiliation{Institut Laue-Langevin, 6 rue Jules Horowitz, BP 156, 38042 Grenoble Cedex 9, France}

\author{Thomas~Prokscha}
\affiliation{Laboratory for Muon-Spin Spectroscopy, Paul Scherrer Institut, CH-5232 Villigen, Switzerland}

\author{Sean~Langridge}
\email{sean.langridge@stfc.ac.uk}
\affiliation{ISIS Facility, STFC Rutherford Appleton Laboratory, Harwell Science and Innovation Campus, Oxon, OX11 0QX, United Kingdom}

\date{\today}


\begin{abstract}
Here we study the effect of La doping in EuO thin films using SQUID magnetometry, muon spin rotation ($\mu$SR), polarized neutron reflectivity (PNR), and density functional theory (DFT). The $\mu$SR data shows that the La$_{0.15}$Eu$_{0.85}$O is homogeneously magnetically ordered up to its elevated $T_{\rm C}$. It is concluded that bound magnetic polaron behavior does not explain the increase in $T_{\rm C}$ and an RKKY-like interaction is consistent with the $\mu$SR data.
The estimation of the magnetic moment by DFT simulations concurs with the results obtained by PNR, showing a reduction of the magnetic moment per La$_{x}$Eu$_{1-x}$O %molecular
for increasing lanthanum doping. This reduction of the magnetic moment is explained by the reduction of the number of Eu-4$f$ electrons present in all the magnetic interactions in EuO films. Finally, we show that an upwards shift of the Fermi energy with La or Gd doping gives rise to half-metallicity for doping levels as high as 3.2~\%.

\end{abstract}
\maketitle
\section{Introduction}
Ferromagnetic semiconductors are attracting great interest for spintronic applications~\cite{Ohno2014, schmehl2007, Dunsiger2010}. However, their technological impact is limited by their low Curie temperature ($T_{\rm C}$). Therefore, it is paramount to find strategies to increase their $T_{\rm C}$ while maintaining their semiconductor-like character.
In the group of ferromagnetic semiconductors, EuO with a $T_{\rm C}$ of 69~K has long been considered a archetypal model system since the exchange interaction is well approximated by the Heisenberg model \cite{Passell1976,Monteiro2013, Lampen2013}. EuO has a magnetic moment of $\sim 7~\mu_{\rm B}$ per Eu$^{2+}$ atom \cite{Barbagallo2010} and a band gap of 1.12 eV at room temperature (RT) \cite{Guntherodt1971}. One of the most remarkable properties of EuO is the Zeeman splitting of the Eu-$5d$ states below 69~K due to the influence of the Eu-$4f$ states, causing a conduction band splitting of about 0.6~eV~\cite{Santos2008}. This high spin splitting energy when coupled with low \textit{n}--doping was proven to transform EuO into a half-metal system \cite{Moodera2007, schmehl2007}. Furthermore, the high localization of the Eu-4$f$ electrons, which mediate the magnetic interactions, renders it  convenient to study the exchange interaction phenomena.

There are several ways of increasing the $T_{\rm C}$ of EuO. The first is by altering the stoichiometry of EuO, e.g. introducing oxygen deficiency into EuO as we have recently shown \cite{Monteiro2013, Barbagallo2010, Barbagallo2011}.
The second is by applying strain \cite{Melville2013}, consequently changing the lattice parameter of the EuO system. However, first principles calculations indicate that the band gap closes at 5~\% of isotropic stress and 6~\% biaxial strain \cite{ingle2008}. The third and most promising mechanism to increase the $T_{\rm C}$ is by doping EuO with other elements. In the group of the lanthanide dopants gadolinium has shown to increase the $T_{\rm C}$ to as much as 170~K for a 4~\% doping concentration \cite{Ott2006, Jutong2015}.

It was claimed recently by Miyazaki \textit{et al.} \cite{Miyazaki2010} that a lanthanum doping concentration of 10~\% increased $T_{\rm C}$ to 200~K, while previous work published by Schmehl \textit{et al.} \cite{schmehl2007} reported a maximum increase of the $T_{\rm C}$ to 118~K for 1~\% lanthanum doping. The latter authors also stated that no further increase in $T_{\rm C}$ occurred for higher doping levels. Previous measurements have also showed a $T_{\rm C}$ of 104~K for a 2~\% La doping sample \cite{Vigren1978}.
This elevation of $T_{\rm C}$ is believed to be caused by the donation of one 5$d$ electron by the La atom into the conduction band strengthening thereby the magnetic interaction between the Eu-Eu next-nearest neighbor atoms ~\cite{AMauger1986, Sattler1972, Vigren1978}.

In a previous publication we studied the effect of \textit{n}--type doping via oxygen deficiency on EuO thin films using $\mu$SR~\cite{Monteiro2013}. We have shown that the results for highly oxygen deficient films were incompatible with the bound magnetic polarons (BMP) \cite{Liu2012} theory and that the RKKY-like interaction is the only known model compatible with our data \cite{Mauger1977}.

Here we study the effect of La doping in EuO thin films, in particular how the doping affects its magnetic moment and how the magnetic order persists above 69 K. This was done by probing the underlying physics by $\mu$SR, PNR and SQUID magnetometry measurements. We have also used DFT to draw a theoretical picture of how the La doping changes the band structure. We also argue theoretically that the increase in doping changes the half-metallic behavior of the films.

The paper is organized as follows: we provide a description of the sample deposition and magnetometry (Section~\ref{section:charecterization}), then use a local probe to address the exchange mechanism (Section~\ref{spacial}). In the second half of the paper we use density functional theory (Section~\ref{dft}) and supporting experimental data (Section~\ref{dft1}) to describe the half-metallicity in La$_{x}$Eu$_{1-x}$O and finish with some conclusions on the effect of doping (Section~\ref{conclude}).

\section{Sample deposition and magnetometry} \label{section:charecterization}
The polycrystalline samples were deposited at 400~$^{\circ}$C using a CEVP magnetron sputtering system with a base pressure of 5$\times$10$^{-9}$~Torr. Co-deposition was performed using three targets: a 99.99~\% pure Eu$_2$O$_3$, a 99.9~\% pure Eu and a 99.9~\% pure La target. The deposition was performed by fixing the Eu$_2$O$_3$ flux and by keeping the total flux from the Eu and La constant. The La doping was controlled by changing the relative flux between Eu and La. The growth was performed in an Ar$^+$ plasma at a pressure of 2~mTorr with a flow rate of 14~sccm. The substrates used were Si(001). A SiO$_2$ layer was deposited between the substrate and the La$_x$Eu$_{1-x}$O film and another on the top as a capping layer to prevent further oxidation of the film. We deposited 10~\%, 15~\% and 20~\% La doped EuO samples.

The SQUID magnetometry measurements were performed in a Quantum Design MPMS 2 system. Each sample was cooled in a field of 50 Oe to 5 K followed by the $M(H)$ measurements  at increasing temperatures. Finally each sample was cooled at zero field and the temperature dependent measurements where performed at different fields. 

\begin{figure}
\includegraphics[width=10cm]{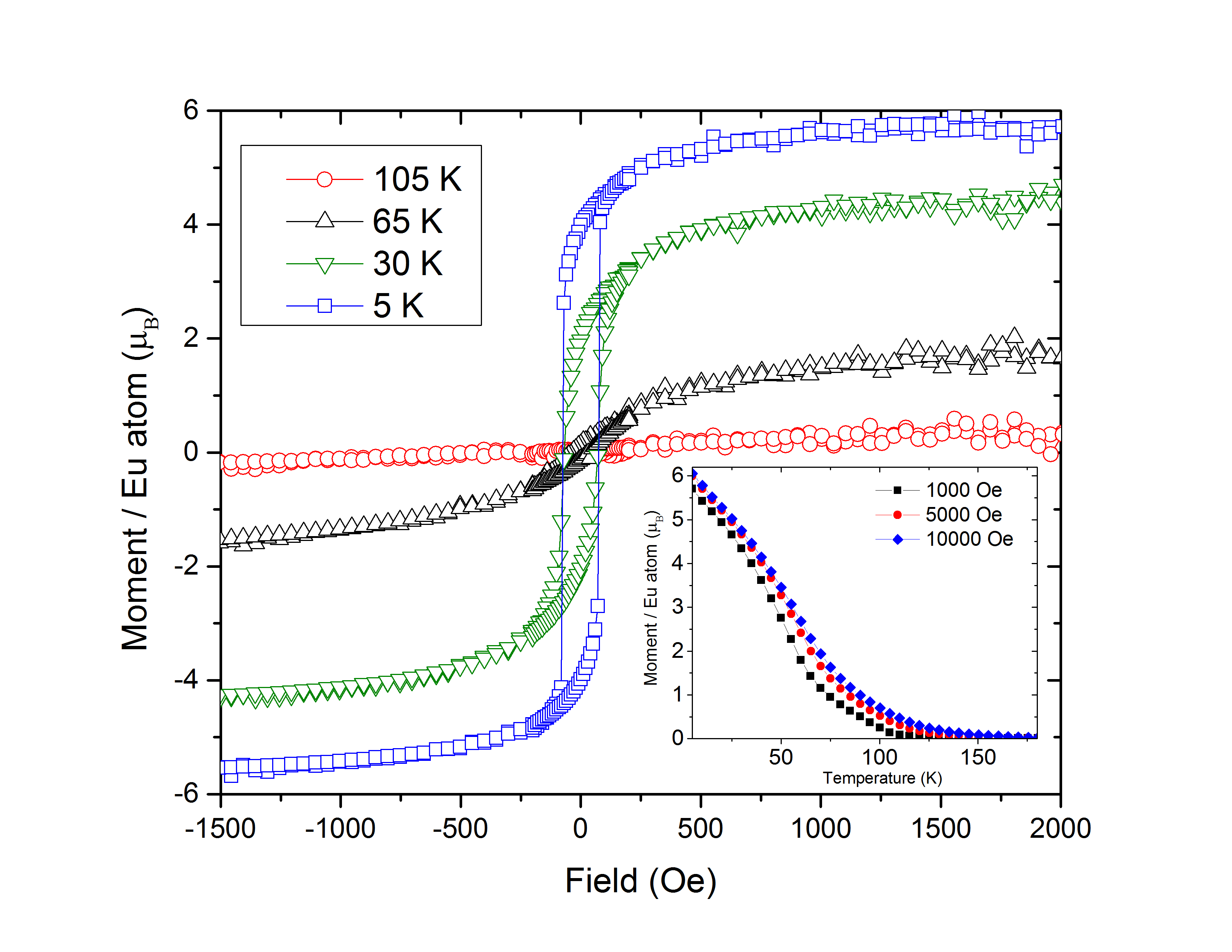}
\caption{Temperature dependent SQUID data for the 15~\% La doped EuO sample. The inset shows the temperature dependence of the magnetic moment. The moment was normalized by the PNR results at 5~K.}
\label{SQUID1}
\end{figure}

\begin{figure}
\includegraphics[width=10cm]{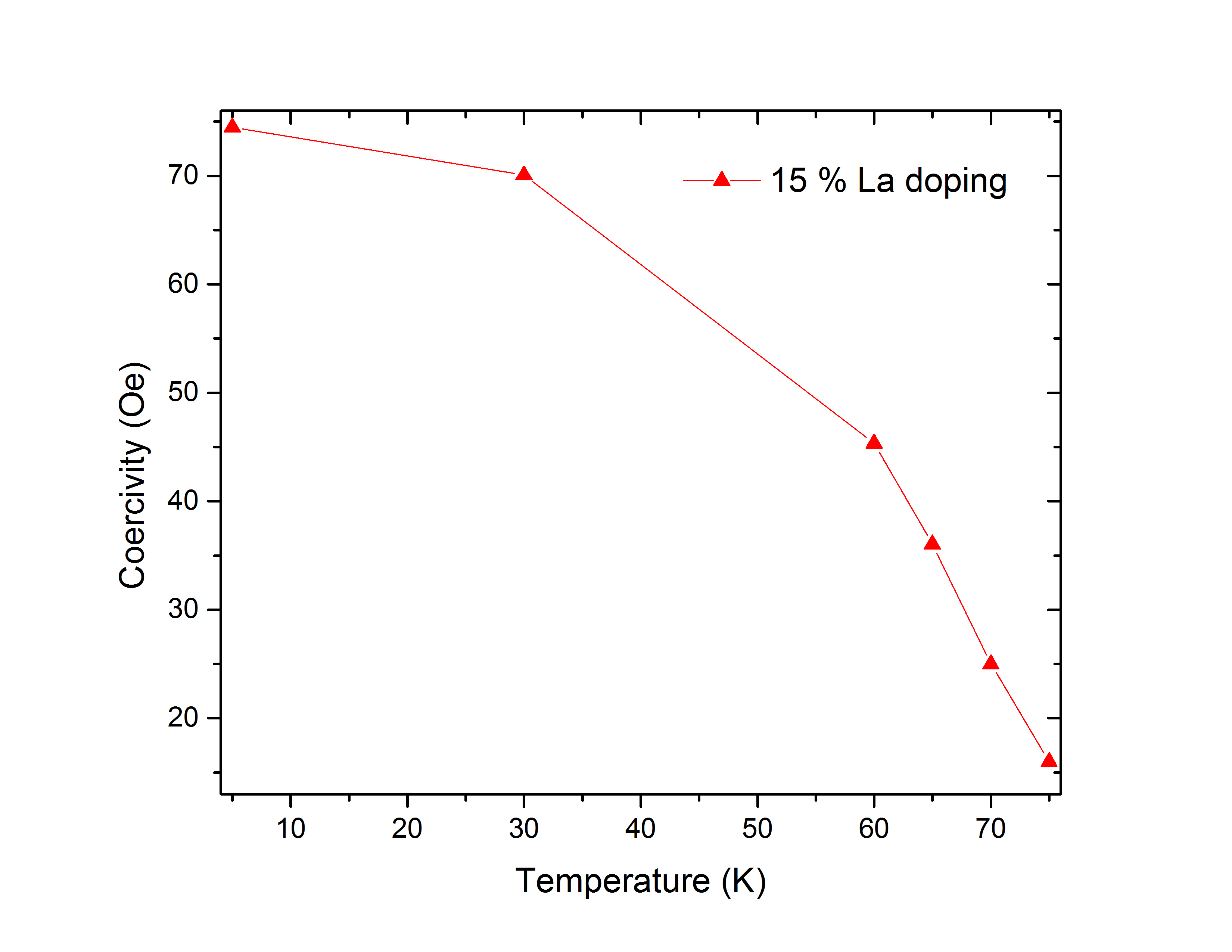}
\caption{Coercivity as a function of temperature for the 15~$\%$ La doped EuO sample.}
\label{coerse15}
\end{figure}

Figure \ref{SQUID1} shows the SQUID data for the 15~\% La doped EuO sample. The hysteresis curve shows a coercive field of 75~Oe compared to 54~Oe for the pristine EuO sample reported in Ref.~\onlinecite{Monteiro2013}. The inset shows the bulk moment obtained at different saturation fields. The inset of Figure \ref{SQUID1} also shows the development of a clear ferromagnetic phase below 96 K, with a paramagnetic response above this temperature.The decrease of the coercivity with temperature is shown in Figure \ref{coerse15}.

Similar measurements of the 10~\% and 20~\% samples find a $T_{\rm C}$ of about 94 K and 84 K, respectively. The lower value for the 20~\% La doped sample is due to disorder and the formation of impurities.  The paramagnetic susceptibility is well described by the Curie-Weiss law and we used this to determine $T_{\rm C}$.

 \section{Magnetic spacial homogeneity}\label{spacial}

\begin{figure}
\includegraphics[width=9cm]{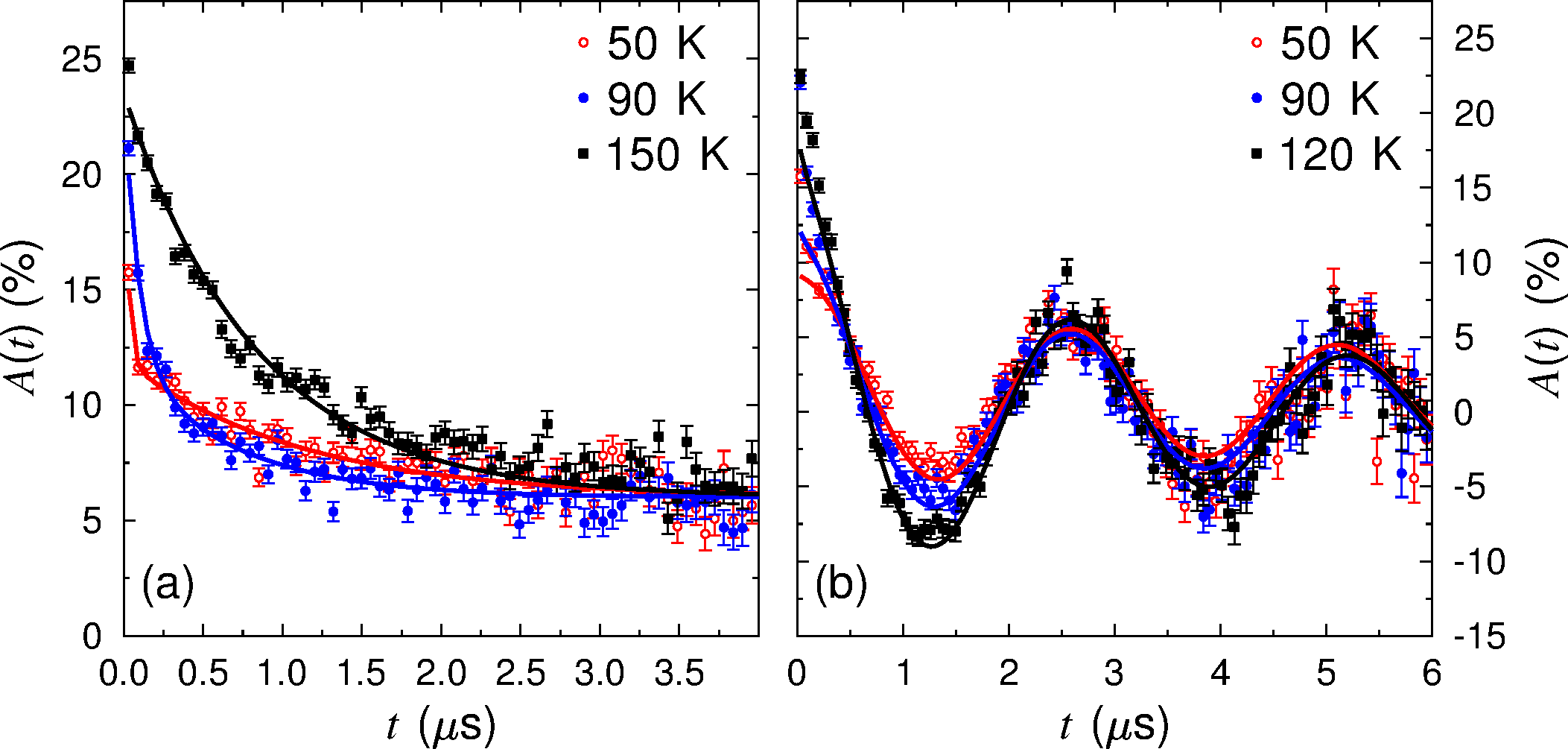}\\
\caption{Raw muon data for the La$_{0.15}$Eu$_{0.85}$O sample:
(a) Zero-field measurements with the fits to Eq.~\ref{muZF}.
(b) Weak transverse field measurements with the fits to Eq.~\ref{muwTF}.
}
\label{mudata}
\end{figure}

The elevated Curie temperature of the EuO system has attracted considerable attention with several models being invoked to describe this behavior~\cite{Mauger1977,Liu2012}.  Low energy muon measurements have been shown to be a highly effective local probe of the magnetism of such systems~\cite{Monteiro2013}. Our measurements were performed at the Paul Scherrer Institute, Switzerland, in zero applied field (ZF) and weak transverse field (wTF = 28.2\,G) configurations, where muons were implanted into the film at various energies between 6 and 14~keV. In $\mu$SR experiments \cite{Blundell2010} spin polarized positive muons are implanted into the sample, where they stop rapidly at interstitial sites of high electron density, and their spin direction evolves in the magnetic field at their stopping site. Each implanted muon decays with a lifetime of 2.2 $\mu$s emitting a positron preferentially in the direction of its spin at the time of decay. The evolution of their spin polarization as a function of time is determined by measuring the direction of the emitted positrons. In low-energy muon experiments~\cite{pt3,pt1,pt2}, accelerating electric fields are used to control the implantation depth of the muons after they have passed through a cryogenic moderator. Example data are shown in Fig.~\ref{mudata}. The lack of clear oscillations below $T_{\rm C}$ in ZF, compared with the data for bulk EuO obtained by Blundell~{\em et al.}~\cite{Blundell2010}, may be due to small variations in the internal fields caused by a distribution of grain boundaries and defects in the film. This was also evident in our previous measurements of EuO$_{1-x}$~\cite{Monteiro2013}. 

\begin{figure}
 \includegraphics[width=8cm]{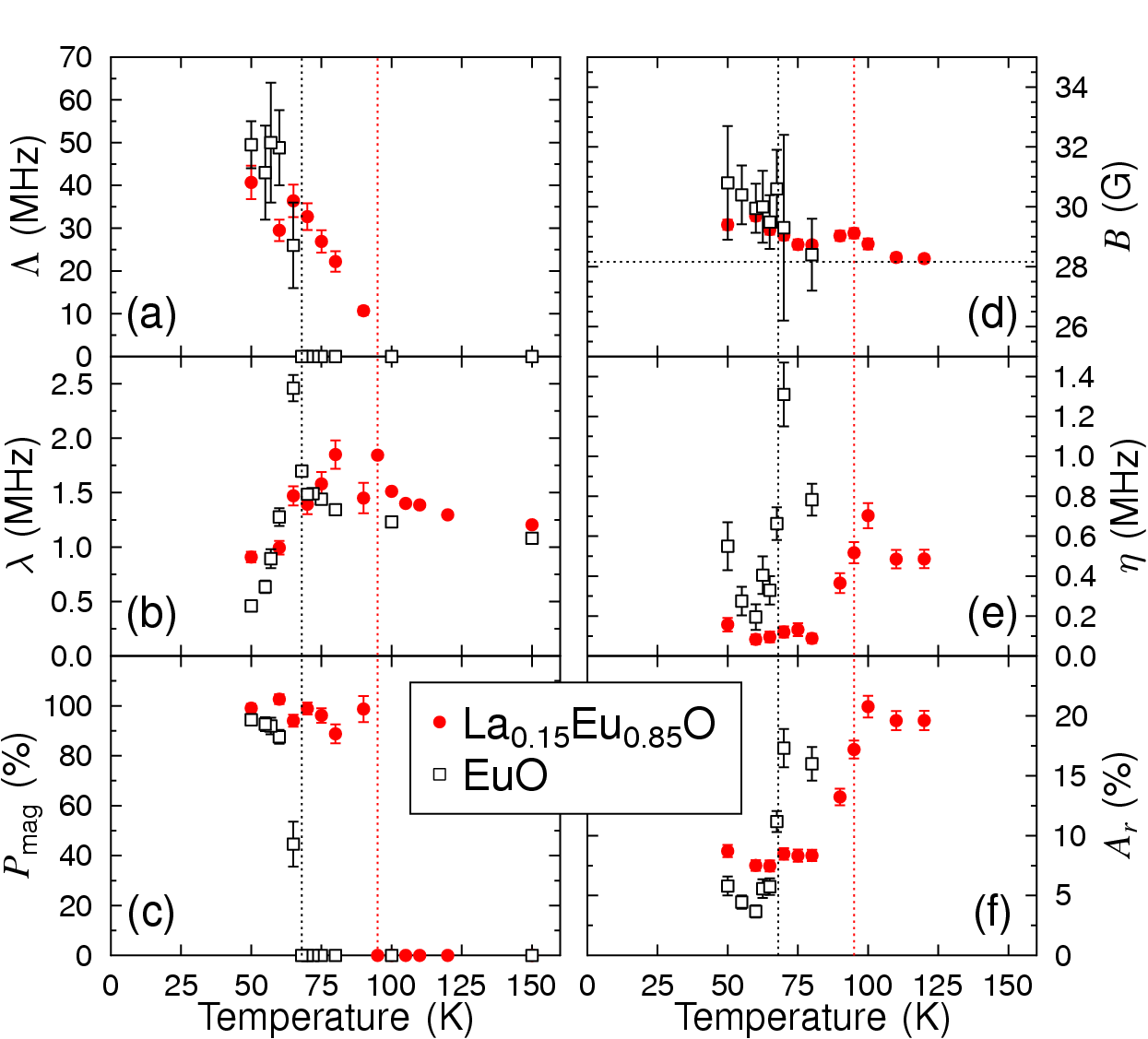}\\
\caption{Fitted parameters for the zero field muon data (a-c) and weak transverse field data (d-f):
(a) Fast relaxation rate, $\Lambda$.
(b) Slow relaxation rate, $\lambda$.
(c) Static magnetic volume fraction, $P_{\rm mag}$.
(d) Magnetic field, $B$, experienced by the implanted muons.
(e) Relaxation rate, $\eta$. (Square-root exponential for La$_{0.15}$Eu$_{0.85}$O and simple exponential for EuO.)
(f) Relaxing asymmetry, $A_r$.
The vertical dotted lines denote the Curie temperatures for EuO (68\,K) and La$_{0.15}$Eu$_{0.85}$O (96\,K), with data for EuO taken from Ref.~\onlinecite{Monteiro2013}.
}
\label{muparameters}
\end{figure}

We described the ZF data using the function~\cite{Monteiro2013}:
\begin{equation}
A(t) = A_1 \exp(-\Lambda t) + A_2 \exp(-\lambda t) +A_{\rm bg},
\label{muZF}
\end{equation}
where the sum of $A_1$ and $A_2$ (17.5~\%), and the background contribution $A_{\rm bg}$ (6~\%) are fixed by the geometry of the spectrometer and sample. The first term, describing the fast relaxation $\Lambda$, is found to be zero above $T_{\rm C}$ and represents an incoherent precession of muons about fields perpendicular to their spin polarization. This relaxation rate $\Lambda$, shown in Figure \ref{muparameters}~(a), will therefore vary with the size of those fields and gives us some insight into the magnetic order parameter, albeit far more limited than a well-defined oscillation frequency. This change in $\Lambda$ exhibits a similar temperature dependence to the magnetization recorded in our bulk measurements, fully consistent with following a ferromagnetic order parameter by both techniques. The second term describes a slow relaxation of muon spins, $\lambda$, which below $T_{\rm C}$ we can attribute to fluctuations of magnetic fields parallel to the muon spin polarization and above $T_{\rm C}$ is caused by paramagnetic spin fluctuations. Figure~\ref{muparameters}(b) shows that there is a small peak in $\lambda$ around $T_{\rm C}$ in La$_{0.15}$Eu$_{0.85}$O that is less pronounced than that in pristine EuO~\cite{Monteiro2013}.

Below $T_{\rm C}$ we note that $A_1 / A_2 \sim 2$, which is the value anticipated for a polycrystalline ordered magnet. Using this we can estimate the fraction of the probed sample volume entering a magnetically ordered state at each temperature as $P_{\rm mag} = 1.5 \times A_1 / (A_1 +A_2)$. In Figure.~\ref{muparameters}~(c) the static magnetic volume $P_{\rm mag}$ is shown as a function of temperature. These data shows that quasistatic magnetic fields develop through the whole sample volume, within the $\sim 5$~\% measurement resolution, at $T_{\rm C}$.

The wTF measurements at low-temperature show a fast relaxation and a slowly-relaxing oscillation, which can be attributed respectively to muons experiencing the large spontaneous fields and the weak applied field. This gives us another probe of the volume of the sample in which the muons are implanted that is magnetically ordered. To simplify the determination of this volume fraction we fitted the data omitting the first $0.25~\mu$s to the function:
\begin{equation}
A(t) = A_r \exp(-\sqrt{\eta t})\cos(\gamma_{\mu} B + \phi).
\label{muwTF}
\end{equation}
This describes the slow relaxing precession of muons not experiencing large internal magnetic fields within the magnetic volume of the sample. The parameters $\eta$ and $B$ are shown in Fig.~\ref{muparameters}~(d) and (e), respectively. The behavior of all three parameters is broadly similar to the equivalent parameters for EuO, except that transverse field relaxation here takes a square root exponential form, rather than the simple exponential observed for EuO$_{1-x}$. This difference can be explained by the presence of an inhomogeneous distribution of fields at muon stopping sites due to non-magnetic site dilution by the La doping. The amplitude $A_r$ [Fig.~\ref{muparameters}(f)] drops on entering the magnetically ordered phase, with the remaining amplitude being consistent with the background determined for the ZF measurements. 

Both ZF and wTF data show that the sample suffers a sharp transition from a fully polarized state to a paramagnetic one. This behavior is similar to what we previously obtained for EuO$_{1-x}$  \cite{Monteiro2013}. Therefore, we conclude that BMP could not be present in La$_{0.15}$Eu$_{0.85}$O and the increase of $T_{\rm C}$ is a consequence of the existence of the RKKY-like interaction between the $4f$ and the $5d$ states.

\section{Half-metallicity in L$\textbf{a} $$_{x}$E$\textbf{u}$$_{1-x}$O} \label{dft}
Of central interest in the EuO system is the half-metallic behavior below the Curie temperature. To investigate this as a function of La doping density functional theory calculations were performed using the CASTEP plane-wave pseudopotential package \cite{Clark2004}. As noted in our previous work \cite{Barbagallo2010}, bulk EuO is poorly described by standard DFT employing local or semi-local functionals, which describe it as a paramagnetic metal instead of a  ferromagnetic insulator. We use instead the DFT+U formalism \cite{Dudarev1998}, which augments the Local Spin Density Approximation (LSDA) with a Hubbard model description penalizing non-integer occupation of the Eu 4$f$ orbitals. As in the previous work, we used $U=7.3$~eV as it accurately reproduces both the experimental bandgap and lattice parameter of bulk EuO. Simulations were performed with a plane-wave basis with a cutoff energy of 600~eV, and a Monkhorst-Pack \textit{k}-point mesh with a minimum spacing of 0.1~\AA$^{-1}$.

We first constructed a 64-atom supercell comprising $2\times 2 \times 2$ copies of the 8-atom conventional cubic cell. Within this cell the calculated magnetic moment per Eu atom is exactly 7.0~$\mu_{\rm B}$. La doping was then introduced by substituting a chosen number of Eu atoms with La atoms and relaxing the resulting geometry.

\begin{figure}
\includegraphics[width=0.38\textwidth]{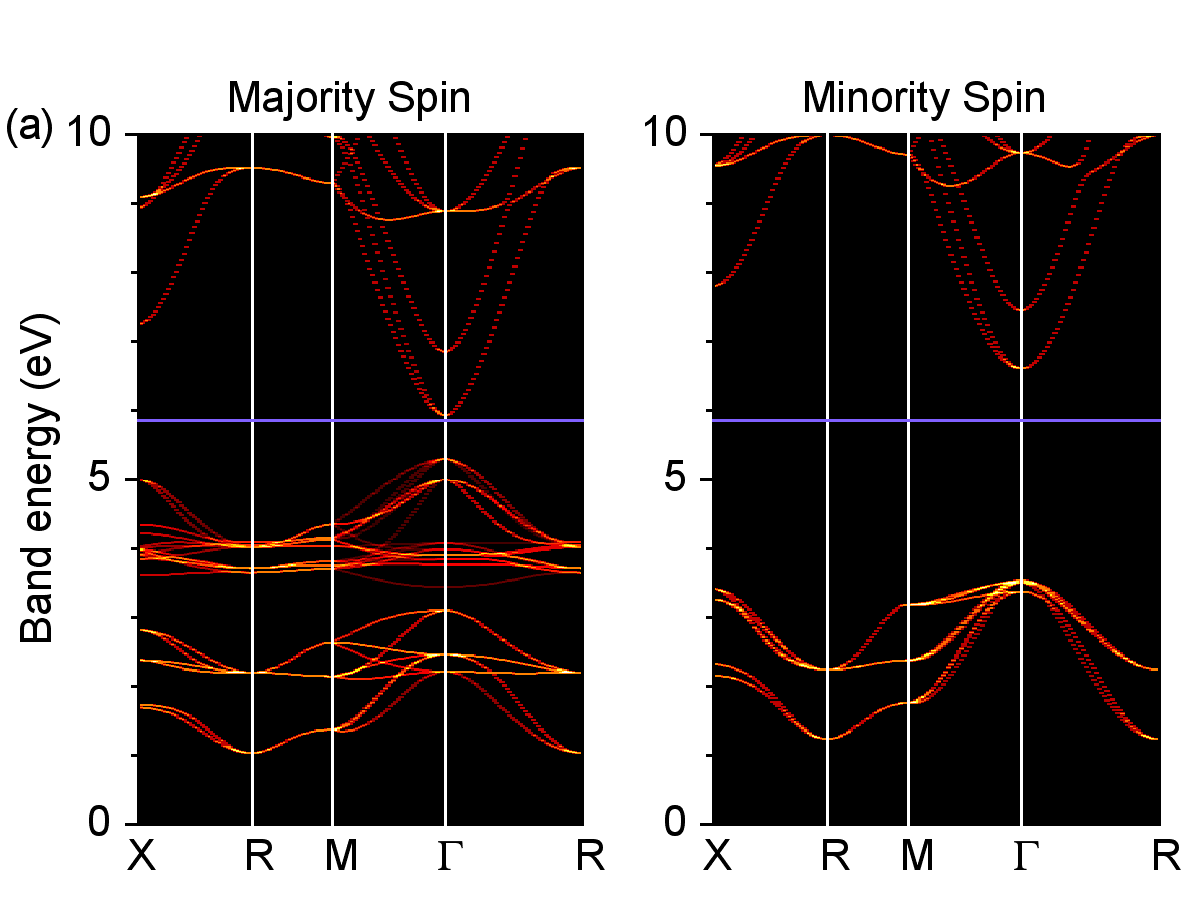}\\
\includegraphics[width=0.38\textwidth]{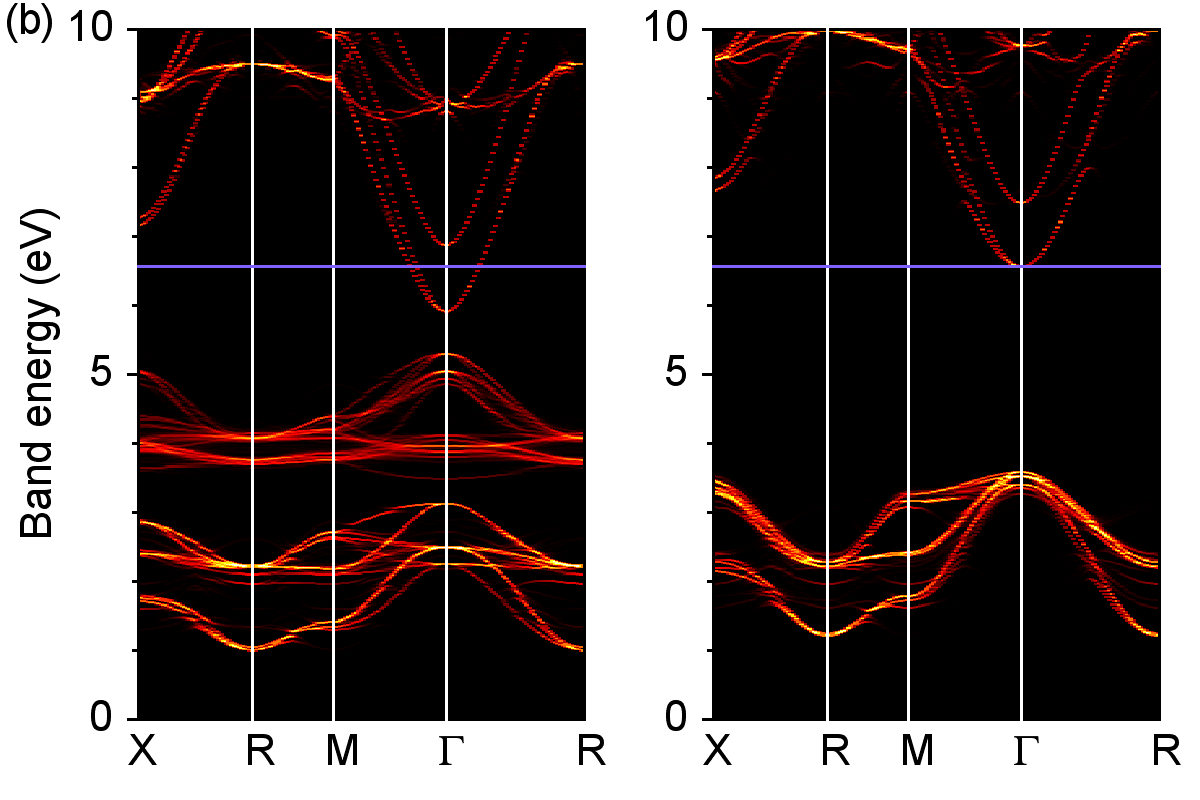}\\
\includegraphics[width=0.38\textwidth]{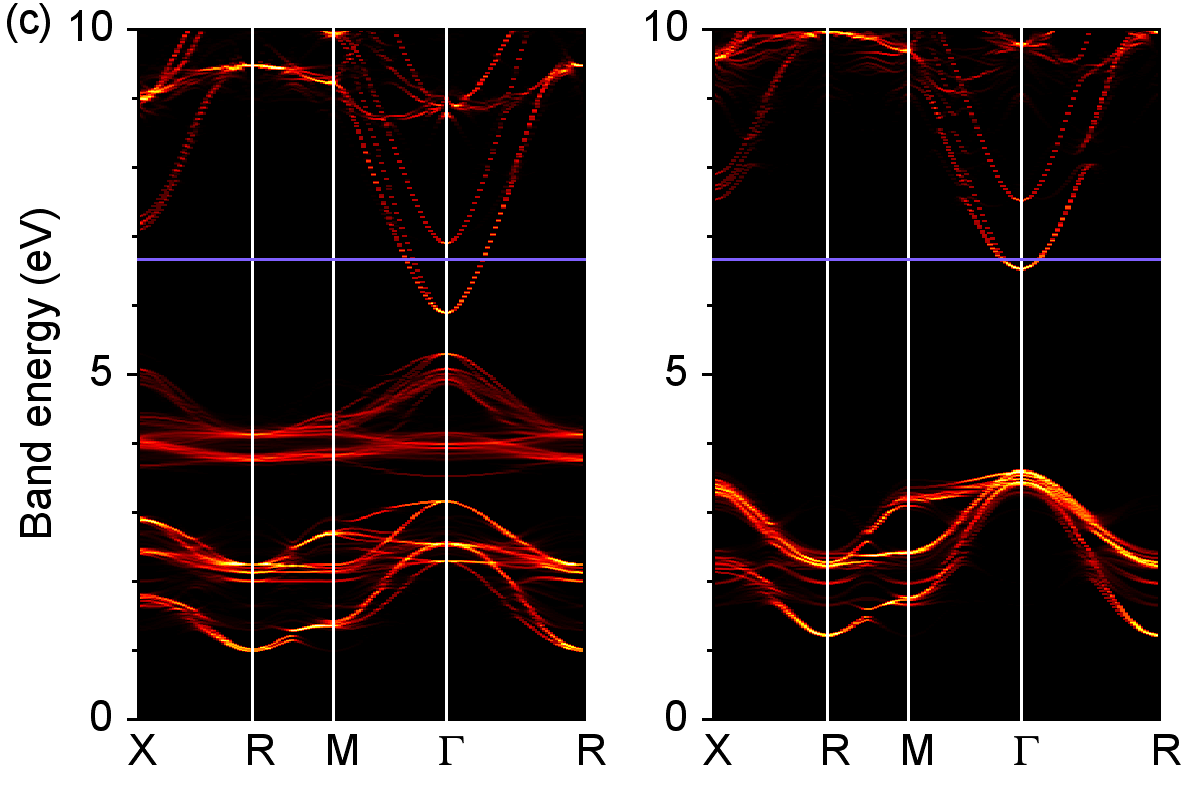}\\
\includegraphics[width=0.38\textwidth]{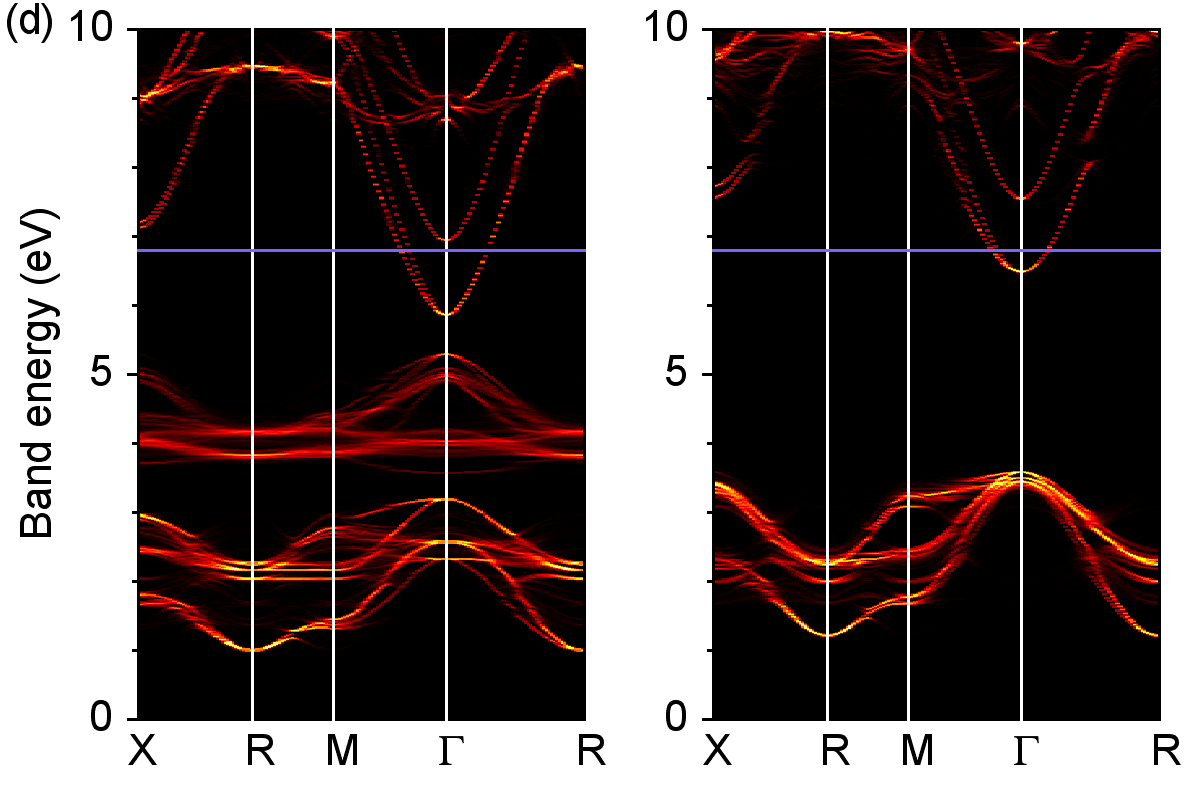}\\
\caption{Effective band structure for a 64-atom EuO supercell: (a) Pristine EuO. (b) 1 La substitutional atom (3.125$\%$). (c) 2 La substitutional atoms (6.25$\%$). (d) 3 La substitutional atoms (9.375$\%$).The La substitutional dopants on Eu sites are projected onto the 8-atom EuO conventional cell. The left panel corresponds the majority-spin states and the right the minority-spin states. The blue line is the Fermi level.}
\label{fig:EBS}
\end{figure}
Since La does not itself have occupied $4f$ states, each La substitution might be expected to reduce the overall magnetic moment by 7.0~$\mu_{\rm B}$. However, in fact the magnitude of the reduction will be less than this, as the La acts as an electron donor to the 5$d$ states comprising the conduction band. The lowest-lying conduction band (at $\Gamma$) is spin-split in the simulations by 0.7~eV due to the effective field from the aligned Eu $4f$ moments. Therefore the electrons occupying the conduction band are at low concentration fully aligned with the magnetic moment of the $4f$ electrons. As the concentration increases, however, the minority-spin channel begins to be filled as well, and the contribution per conduction electron falls. At this point the half-metallicity is lost as the Fermi level crosses both majority and minority spin states Fig.~\ref{fig:EBS}~(c).

Figure~\ref{fig:EBS} shows the effective band structure (EBS) of the La-doped supercell calculations, projected into the Brillouin zone of the conventional cubic unit cell of perfect EuO. We use the method of Popescu and Zunger \cite{Popescu2012} for the EBS plots, as implemented by Brommer {\em et al.}~\cite{Brommer2014}. As the La-doping fraction $x$ is increased, we see the Fermi level rise further into the conduction band around $\Gamma$. Simultaneously, the $4f$ bands become less distinct, as the La dopant atoms both disrupt the periodicity of the system and reduce its overall magnetic moment. At high $x$, we see evidence of subbands forming in many regions of the plot, but this is due to the artificial periodicity of the 64-atom cell and would not be observed so clearly for truly random distributions of defects in larger cells.

Mulliken population analysis using pseudo-atomic orbitals as projectors indicates that the magnetic moment is not strongly localized on La sites. These only have a net spin in the range 0.1-0.2~$\mu_{\rm B}$ depending on the number of La atoms present in the cell. Instead, the net magnetic moment is distributed over the whole cell, indicating that it occupies rather delocalized $5d$ orbitals.

\begin{figure}
\includegraphics[width=9cm]{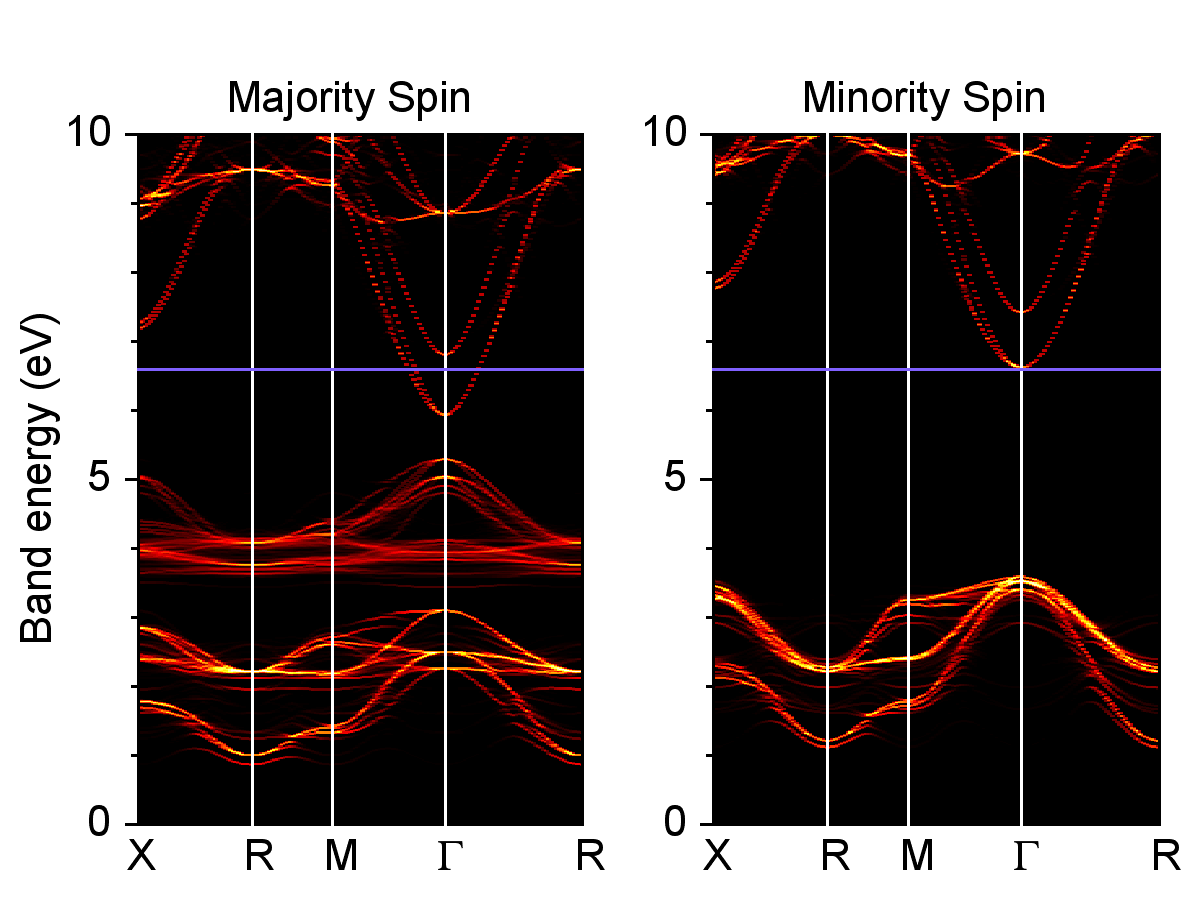}
\caption{Effective band structure for a 64-atom EuO supercell with 1 Gd substitutional atom (3.125$\%$). The substitutional dopants on Eu sites are projected onto the 8-atom EuO conventional cell. The left panel corresponds the majority-spin states and the right the minority-spin states. The blue line is the Fermi level.}
\label{Gd}
\end{figure}

Given the high ordering temperature with Gd doping it is also opportune to compare the effects of La and Gd doping in order to better understand the phenomena of electron doping. We have therefore performed 
DFT calculations substituting a number of Eu atoms for Gd, and we show effective band structures from these simulations in Fig. \ref{Gd}. In comparison to the La-doped simulations, doping with Gd shows a overall similar behavior in relation to the position of the Fermi level with respect to the majority and minority spin bands, but the $4f$ density of states remains constant for increasing doping levels as there is no decrease in $4f$ occupancy per unit cell. Therefore, appropriate levels of Gd doping can also be expected to result in a half-metallic material.

\section{Variation of the magnetic moment with L$\textbf{a}$ doping}\label{dft1}

\begin{figure}
\includegraphics[width=9cm]{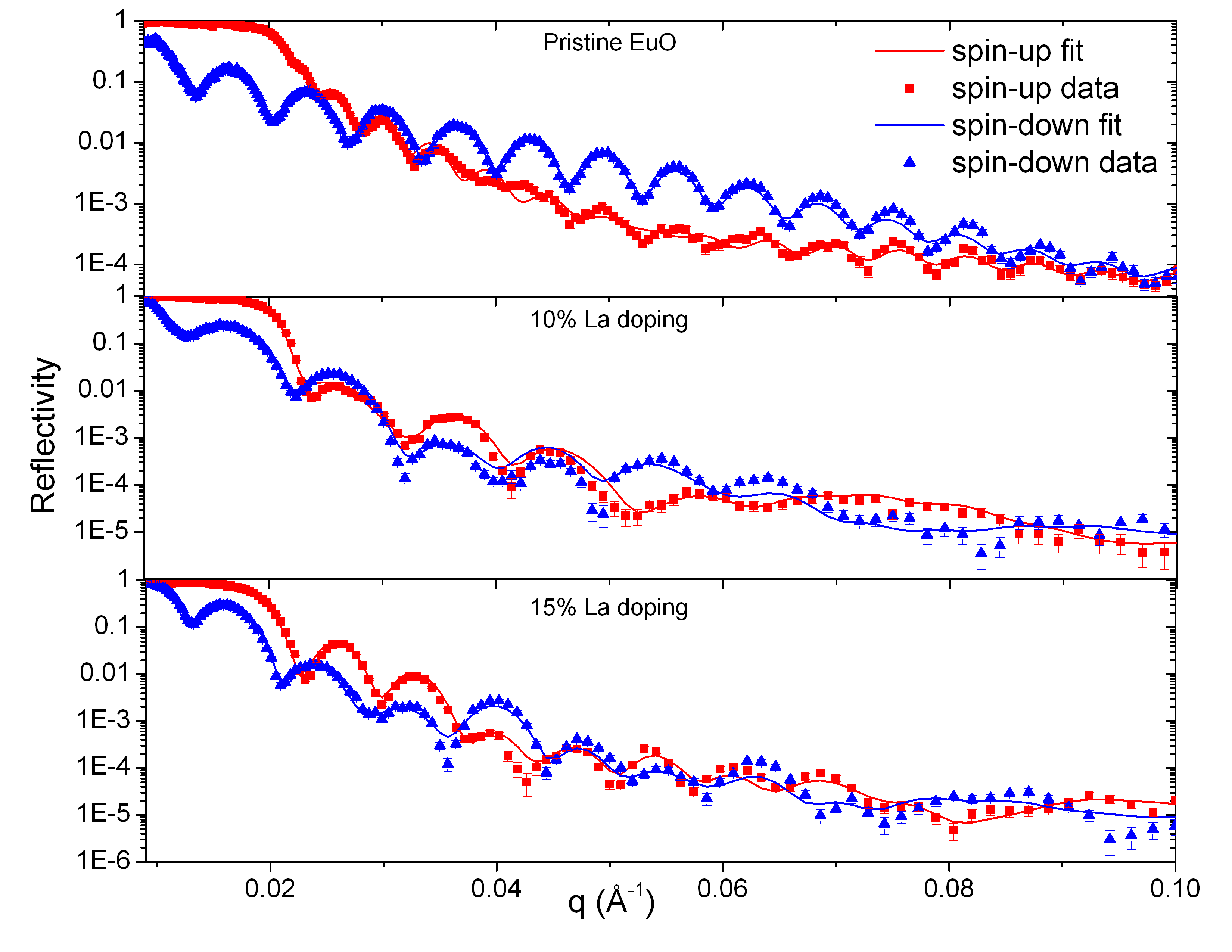}
\caption{PNR data and analysis for the pristine EuO, 10~\% and 15~\% La doped EuO samples at 5~K. The spin-up and spin-down data is shown in red and blue, respectively.}
\label{PNR15}
\end{figure}

To investigate the critical effect of the La doping in EuO thin films one needs to understand the impact it has on the overall magnetic moment and also if any substantial changes occur in the bandstructure. The experimental determination of the magnetization of the La$_x$Eu$_{1-x}$O films was done using low temperature PNR 
on the neutron reflectometer CRISP at ISIS, UK, following preliminary measurements on beamline D17 at the ILL, France. The data analysis was performed by using the GenX~\cite{Bjorck2007} and XPolly fitting softwares.
Figure \ref{PNR15} shows the PNR reflectivity obtained for the three measured La$_{x}$Eu$_{1-x}$O samples at 5 K, showing that the 15~\% La doping sample has a reduced magnetic moment of 5.7 $\mu_{\rm B}$ per La$_{0.15}$Eu$_{0.85}$O formula unit in comparison with the 6.98 $\mu_{\rm B}$ per stoichiometric EuO formula unit.

\begin{figure}
\includegraphics[width=8.5cm]{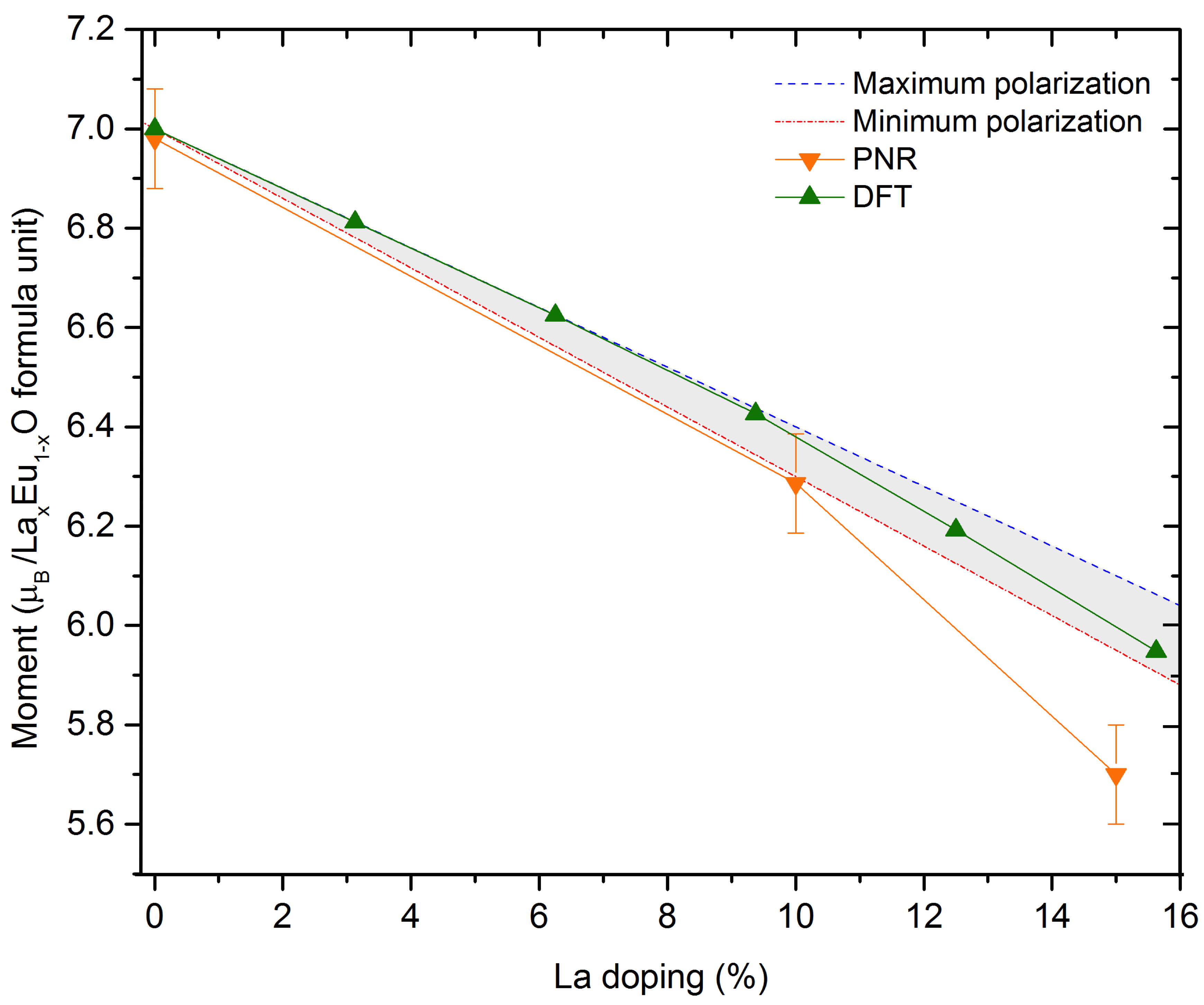}\\
\caption{Comparison between the DFT calculations and the PNR analysis. The blue-dashed and red-dotted-dashed lines corresponds to the maximum and minimum polarization coming from the doping electrons, respectively.}
\label{fig:comparison}
\end{figure}

Figure \ref{fig:comparison} compares the magnetic moment calculated using DFT and measured by PNR. The figure shows a good agreement between the theory and the experiment at 10~\% doping. The 15~\% lanthanum doped EuO sample has a magnetic moment slightly reduced compared with the DFT calculation and we believe this is owing to the fact that the random distribution of the La atoms creates disorder that weakens the magnetic interactions. Another explanation is the presence of a small percentage of impurities lowering the overall magnetic moment measured by PNR. The figure also shows the calculated maximum and minimum possible polarization of the conduction band, in a model which assumes that at concentration $x$ of La substituting Eu, the moment of the $4f$ electrons will be given by $7(1-x)$ $\mu_B$. The maximum polarization curve assumes a further $x$ $\mu_B$ contribution from fully-polarized electrons in the conduction band, while the minimum polarization curve assumes they are fully unpolarized. As explained before, the doping electrons populate only the majority-spin states at low doping levels. However as the doping increases, the Fermi energy is shifted upwards, crossing the minimum of the minority-spin conduction bands. Beyond this point both minority and majority spin states will be occupied and the overall polarization drops. This can be seen in Figure \ref{fig:comparison}: the DFT and maximum polarization curves start at similar values, and at about $x=0.0625$ they begin to diverge. This divergence is attributed the loss of the half-metallicity that occurs at $x=0.03125$.

\section{Conclusions}\label{conclude}

In this paper we have shown that the La doping acts like an \textit{n}--dopant increasing the density of carriers in the 5$d$ conduction band and strengthening the $4f$-$5d$ (Eu-Eu) interaction. The $\mu$SR experiments show that the electron doping in La$_{x}$Eu$_{1-x}$O thin films has a similar effect on the magnetic properties to that which was found for EuO$_{1-x}$. 
 One of the indicators is the fact that P$_{mag}$ develops an abrupt transition at the elevated $T_{\rm C}$ (96 K).
  This is a crucial indicator that there is no magnetic phase separation by the presence of bound magnetic polarons. It also indicates that no significant clustering of La is occurring, as this would broaden the transition. In the absence of bound magnetic polarons the electrons populate the 5$d$ band and the RKKY-like interaction is the one that explains the $T_{\rm C}$ enhancement at high doping levels.

The PNR and DFT show a reduction of the overall magnetic moment of the La$_{x}$Eu$_{1-x}$O formula unit. This is attributed to the replacement of Eu by La atoms reducing the total amount of 4$f$ electrons in the system. The 4$f$ electrons are responsible for carrying the magnetic moment and when removed a dilution of the total magnetic moment in the system occurs. The difference between the PNR and DFT at higher doping levels is ascribed to disorder and impurities in the sample.

The DFT calculations predicts that the band structure remains almost unaffected for increasing doping where the only change is an upwards shift of the Fermi energy. For doping levels below 3.125~\% this change is just enough to make the Fermi energy intersect the majority spin states leaving the minority spin stated unoccupied -- forming a fully spin polarized conduction band. This half-metallic behavior occurs at much higher levels than the ones reported for oxygen deficient EuO, 
where at the same vacancy doping levels the system was already showing both the majority and minority channels occupied. The DFT calculations also show that the range of doping where half-metallicity can be expected in La or Gd doped EuO is limited by the size of the conduction band spin splitting. The presence of half-metallicity with La or Gd doping makes EuO a extraordinary system to study the phenomena of half-metallicity in strongly correlated  systems.

\begin{acknowledgments}
Parts of this work were performed at the Swiss Muon Source, Paul Scherrer Institute, Villigen, Switzerland; ISIS, STFC Rutherford Appleton Laboratory, UK; and the ILL, France. All calculations were carried out using the Darwin Supercomputer of the University of Cambridge HPC Service. We thank Francis Pratt for useful discussions on TRIM.SP calculations. Pedro Monteiro would like to thank the support of EPSRC (UK), and FCT (SFRH/BD/71756/2010), Portugal. NDMH acknowledges the support of the Winton Programme for the Physics of Sustainability.
\end{acknowledgments}

\bibliography{LaEuO}

\end{document}